\documentclass[12pt,preprint]{aastex}
\usepackage{natbib}
\def\lapprox{$_<\atop{^\sim}$}
\def\grapprox{$_>\atop{^\sim}$}
\begin{document}
\title{Continuum and CO/HCO$^+$ Emission from the Disk Around the T~Tauri
Star LkCa~15}
\author{Chunhua~Qi\altaffilmark{1,2},
Jacqueline~E.~Kessler\altaffilmark{3},
David~W.~Koerner\altaffilmark{4},
Anneila~I.~Sargent\altaffilmark{5}, \&
Geoffrey~A.~Blake\altaffilmark{1,3}}

\altaffiltext{1}{Division of Geological \& Planetary Sciences,
California Institute of Technology 150--21, Pasadena, CA 91125,
USA.} \altaffiltext{2}{ Harvard--Smithsonian Center for
Astrophysics, 60 Garden Street, MS 42, Cambridge, MA 02138, USA.}
\altaffiltext{3}{Division of Chemistry \& Chemical Engineering,
California Institute of Technology, Pasadena, CA 91125, USA.}
\altaffiltext{4}{Department of Physics and Astronomy, Northern
Arizona University, P.O. Box 6010, Flagstaff, Arizona 86011, USA.}
\altaffiltext{5}{Division of Physics, Mathematics \& Astronomy,
California Institute of Technology 103--33, Pasadena, CA 91125,
USA.}
\begin{abstract}
We present OVRO Millimeter Array $\lambda = 3.4 - 1.2$ mm dust
continuum and spectral line observations of the accretion disk
encircling the T Tauri star LkCa~15.  The 1.2 mm dust continuum
emission is resolved, and gives a minimum diameter of 190 AU and
an inclination angle of $57 \pm 5^{\circ}$.  There is a noticeable,
but at present poorly constrained, decrease in the continuum
spectral slope with frequency that may result from the coupled
processes of grain growth and dust settling. Imaging of the fairly
intense emission from the lowest rotational transitions of CO,
$^{13}$CO and HCO$^+$ reveals a rotating disk substantially larger
than that observed in the dust continuum. Emission extends to
$\sim$750 AU and the characteristic radius of the disk is
determined to be $\sim$425 AU (HWHM), based on model fits to the CO
velocity field. The measured line ratios demonstrate that the
emission from these species is optically thick, while that from
C$^{18}$O and H$^{13}$CO$^+$ is optically thin, or nearly so. The
disk mass derived from the CO isotopologues with ``typical'' dense
cloud abundances is still nearly two orders of magnitude less than
that inferred from the dust emission, the most probable
explanation being extensive molecular depletion in the cold, dense
disk midplane. Thus, while CO, HCO$^+$ and their isotopologues are
excellent tracers of the disk velocity field, they are not
reliable tracers of the disk mass.

N$_2$H$^+$ 1$\rightarrow$0 emission has also been detected which, along
with HCO$^+$, sets a lower limit to the fractional ionization of 10$^{-8}$
in the near-surface regions of protoplanetary disks. This first
detection of N$_2$H$^+$ in circumstellar disks has also made possible
a determination of the N$_2$/CO ratio ($\sim$ 2) that is at least
an order of magnitude larger than those in the envelopes of young
stellar objects and dense clouds. The large N$_2$/CO ratio indicates
that our observations probe disk layers in which CO is depleted but
some N$_2$ remains in the gas phase.  Such differential depletion
can lead to large variations in the fractional ionization with height
in the outer reaches of circumstellar disks, and may help to explain
the relative nitrogen deficiency observed in comets.

\end{abstract}


\section{Introduction}

LkCa~15, located in the Taurus-Auriga complex at a distance of 140 pc, is
rapidly becoming one of the best studied T Tauri stars at
millimeter wavelengths.  A K5 star with an optical/near-infrared
luminosity of $\sim$0.8 L$_{\odot}$ (\citealp{elias78},
\citealp{neuhaeuser_s95}, \citealp{simon_d01}), LkCa~15 is well
isolated from dark clouds and lacks any detectable reflection
nebulosity in HST WFPC2 images (\citealp{krist_b97}). Its
substantial near- through far-infrared excess is consistent with
the presence of a circumstellar disk (\citealp{strom_s89}) whose
mass and radial extent have been determined by recent spectral
energy distribution fits and millimeter-wave interferometric
observations to be some 0.024 M$_{\odot}$ and 650 AU, respectively
(\citealp{chiang_j01, simon_d01}).

Age estimates for LkCa~15 range from 3 Myr (\citealp{simon_d01})
to 12 Myr (\citealp{thi01}). These time scales are similar to those
suggested for the disappearance of small dust grains in the inner
regions of circumstellar disks (\citealp{skrutskie90,lada99,robberto_m99})
and for the differentiation of meteoritic
material in the early solar nebula (\citealp{alexander_b01, amelin_k02}).

The large size and mass of the disk, combined with the age of LkCa
15, make this an important system for further study since it may
represent an important transitional phase in which viscous disk
spreading and dispersal competes with planetary formation
processes. Further, the outer disk can be studied in detail with
existing (sub)millimeter-wave telescopes.  Single dish
observations have already revealed readily detectable emission
from several molecules (\citealp{vanzadelhoff_v01,thi01}), while
aperture synthesis imaging has the potential to shed light on the
chemical and physical gradients in the disk.

Our initial millimeter-wave observations of LkCa~15 have
concentrated on species that are typically abundant in dense molecular
clouds. Here we present Owens Valley Radio Observatory (OVRO)
millimeter array observations of the disk encircling LkCa~15 in
dust continuum emission and the lowest rotational transitions of
various isotopologues of CO and HCO$^+$. An upper limit to the
emission from the H$^{13}$CO$^+$ ion is noted. N$_2$H$^+$ was also
observed to constrain the fractional ionization and molecular
depletion beyond 100 AU. Due to the limited angular extent of the
disk, small scale gradients were largely smoothed out by the spatial
resolution achieved, and the resultant line intensities should be
considered as an ``average'' over the outer disk.  Nevertheless,
these interferometric observations suffer from substantially less
beam-dilution than those with single telescopes; and this,
combined with the inherent stability of cross-correlation
measurements, provides a powerful means of studying the weak
emission from species such as N$_2$H$^+$ and C$^{18}$O. While
arrays under development will provide substantially improved
sensitivity and spatial resolution, observations at the
{\lapprox}few arcsecond level form the first and necessary step to
understanding depletion and ionization within the disks of T Tauri
stars. Future publications will present more detailed analyses of
these and other data acquired in an extensive millimeter-wave
interferometric spectral line survey of LkCa~15
(\citealp{qi03,kessler03}). In what follows, \S2
includes a brief overview of the observations, while in \S3 we
discuss the results and the important effects of gas-phase
depletion and fractional ionization on the physical and chemical
structure of the disk around LkCa~15. A brief summary of our
results and the role of future observations may be found in \S4.

\section{Observations}

All measurements were made between 1997 October and 2000 April
using the OVRO Millimeter Array at Big Pine, California. The array
consists of six 10.4 meter telescopes with an rms surface
precision of 35 microns. The pointing accuracy is about 4$''$,
except for brief excursions at sunrise and sunset.  Combinations
of five array configurations were used to map LkCa~15, with
antenna spacings ranging from 20 to 400 m east-west and from 25 to
440 m north-south. Cryogenic SIS receivers on each telescope
produced average single sideband system temperatures of 450, 1300,
and 1200 K at the frequencies of the $^{13}$CO 1$\rightarrow$0, CO
1$\rightarrow$0, and CO 2$\rightarrow$1 lines, respectively. The
receivers were tuned for double sideband operation so that both
the upper and lower sidebands could be used for molecular line
studies.

The receiver IF output is sent over fiber optic transmission
lines to a pair of 1 GHz bandwidth analog correlators and a 512
lag digital correlator with four separately configurable modules.
Spectral line and continuum measurements are made simultaneously
and can be carried out in both the $\lambda$ = 3 and 1.3 mm
atmospheric windows with a single local oscillator set up.
During the $^{13}$CO 1$\rightarrow$0 and HCO$^+$ 1$\rightarrow$0
observations of LkCa~15,
the digital correlator was configured into two bands of
64 $\times$ 0.125 MHz channels to provide a velocity
resolution of 0.17 and 0.21 km s$^{-1}$, while the CO 2$\rightarrow$1
measurements were made using 256 $\times$ 0.125 MHz channels for
a velocity resolution of 0.16 km s$^{-1}$.

Calibration of the visibility phases and amplitudes was achieved with
observations of the quasars 0528+134 and 0507+179, typically at intervals
of 20--30 minutes.  Measurements of Uranus,
Neptune, 3C 273 and 3C 454.3 provided an absolute scale for the
calibration of flux densities. All data were phase- and
amplitude-calibrated using the MMA software package, developed
specifically for OVRO (\citealp{scoville_c93}).
Continuum and spectral line maps were generated and CLEANed using the
NRAO AIPS package. The continuum level is not subtracted in obtaining
the line maps because most of the transitions observed here are
optically thick whereas the continuum is optically thin.
Uncertainties in the fluxes and source positions are
estimated to be 20\% and 0.$''$5, respectively. The synthesized beams
vary from 1$''$ at 1.2 mm to 4$''$ at 3.4 mm.

\subsection{Continuum Emission}

Figure~\ref{fig:continuum} shows the spectral energy distribution
of the continuum emission from LkCa~15. A noticeable, but at
present poorly constrained, decrease in the continuum spectral
slope with frequency has been found, as discussed further in \S3.1.
The dust emission is spatially resolved, but only in the highest
resolution OVRO 1.2 mm images. The ($u,v$)-plane fits of a
two-dimensional gaussian to the highest quality visibilities yield
a deconvolved size of $(1.''35\pm0.''12)\times(0.''74\pm0.''06)$
at a position angle (PA) of $66\pm5 ^{\circ}$, which corresponds to
a minimum size of the dust disk of $189\pm17$ AU.  Assuming the
disk is circular, the measured ellipticity indicates an
inclination angle of $57\pm5 ^{\circ}$ (0$^{\circ}$ being the
inclination of a face-on disk). This inclination is somewhat
larger than the value obtained by \citet{duvert_g00},
34$^{\circ}$, but smaller than the value of 66$^{\circ}$ obtained
by \citet{kitamura_m02} with their NMA 2 mm continuum
observations, probably due to differences in the synthesized
beams. Nevertheless, it is in good agreement with both the OVRO
and Plateau de Bure measurements of the disk properties as traced
by CO (\citealp{simon_d01}).

\subsection{CO/HCO$^+$ Observations}

Figure~\ref{fig:co} shows the integrated intensity and
intensity-weighted mean velocity maps of the CO 2$\rightarrow$1
emission toward LkCa~15.  As shown in the left panel, the
molecular gas emission appears much more extended than the dust.
Solid grains no doubt exist at similar radii, but are undetectable
due to the more rapid fall-off in dust emissivity
(\citealp{sargent_b87, lay_c97}). Conversely, the molecular emission from the
innermost region of the disk (R$<$30-50 AU) is sufficiently beam
diluted even at 1-2$''$ resolution that no direct information on
this size scale can be obtained without further improvements in
sensitivity and imaging performance. The circumstellar gas
emission is an effective tracer of the line-of-sight velocity
component of the outer disk via Doppler shifts, as is shown by the
intensity-weighted mean gas velocity map in Figure 2 (right panel).
Molecular gas southwest (SW) of the star is blue-shifted,
whereas that to the northeast (NE) is red-shifted, consistent with
gas orbiting in a disk inclined with respect to the plane of the
sky. With a beam size of $\sim$2$''$ (or 280 AU), the
OVRO CO 2$\rightarrow$1 image of the LkCa~15 disk is well
resolved. A gaussian fit to the velocity-integrated
CO intensity map (Figure~\ref{fig:co}) results in a FWHM of
$(4.''4\pm0.''4)\times(3.''6\pm0.''3)$ at a position angle of
68$^{\circ}$, which corresponds to a radius of {\grapprox}300 AU.

At such moderately large inclinations shear broadening exceeds
thermal broadening in the spectrum, and the local emissivity of
optically thin lines becomes anisotropic (\citealp{horne_m86}). In
this case, the separation of the two peaks in the CO 2$\rightarrow$1
spectrum (2.6 km s$^{-1}$, see Figure~\ref{fig:co_spectra})
becomes 2V$_{kep}$sin$i$, where V$_{kep}$ is the
Doppler shift due to the Keplerian orbital motion, or
$\sqrt{\frac{GM_{*}}{R_{d}}}$, indicating $R_{d}$\grapprox313 AU.
In Figure~\ref{fig:cochan}, observed maps on velocity channels of
width 0.65 km s$^{-1}$ are compared with synthetic maps generated
by a detailed model of a disk in Keplerian rotation. Following
\citet{koerner95}(see also \citealp{mannings_k97}), the model used
a two-component gaussian emissivity distribution with half-maximum
radii taken from the fit to the integrated intensity map in
Figure~\ref{fig:co}. The inclination angle $i$, outer cut-off
radius $R_{d}$, and relative amplitudes of the gaussian components
were varied to match the structure in Figure~\ref{fig:cochan}.
Best fits were found for a central mass $0.9 \pm 0.2~M_{\odot}$
and disk inclination of $i_{disk}=58^{\circ} \pm 10^{\circ}$. The
outer gaussian component has a HWHM of 425 AU and a
cut-off radius of 750 AU. These values are in good agreement with
optical and infrared photometry (\citealp{bouvier_c95}), PdBI CO
observations (\citealp{simon_d01}) and with our resolved 1.2 mm
continuum observations. As can be seen in Figure~\ref{fig:cochan},
synthetic maps produced from this model agree very well with the
CO J=2$\rightarrow$1 spectral line
maps of LkCa~15. The size of the gas disk is thus well determined
both from the image and the spectral data cube, and is much larger
than the dust disk size as determined from continuum emission.
In what follows, important disk physical properties are fixed at the
values derived from the dust (disk mass) and CO (disk size and
orientation) emission.

For robust measurements of the abundance and excitation of CO,
it is necessary to derive observational estimates of line opacities. These
 can be obtained by examining a range of transitions and
isotopologues.  Visibilities from several configurations were
therefore combined to examine the $^{13}$CO, C$^{18}$O, HCO$^+$,
and N$_2$H$^+$ J=1$\rightarrow$0 transitions; results are
presented in Figure~\ref{fig:molecules} and
Figure~\ref{fig:molecules_spectra}. The velocity patterns observed
in both $^{13}$CO and HCO$^+$ are entirely consistent with those
seen in CO, with red-shifted emission to the NE and blue-shifted
emission to the SW. The C$^{18}$O and N$_2$H$^+$ 1$\rightarrow$0
data cubes have lower signal-to-noise values, and so no velocity
analysis was carried out for these transitions. The integrated
line intensities and the corresponding column densities for the
above transitions were derived with the assumptions of local
thermodynamic equilibrium (LTE) and optically thin emission at an
excitation temperature of 30 K as presented in
Table~\ref{tab:mol}. Only upper limits have been obtained for
H$^{13}$CO$^+$.

\section{Discussion}

\subsection{Disk Structure and Grain Properties}

Two of the most important global properties of a disk are its mass
and temperature distribution. At sufficiently short wavelengths,
the entire disk is optically thick, and the density distribution
does not affect the emission. At wavelengths longer than about 300
$\mu$m, however, the thermal emission from circumstellar disks
becomes optically thin. Thus, continuum observations at millimeter
wavelengths provide an excellent way to measure disk masses
directly via the equation:
\begin{equation}\label{eq:mass}
F_{\nu }=\int \, I_{\nu }d\Omega =\int \, S_{v}\tau _{v}d\Omega \approx \int \frac{2kT}{\lambda ^{2}} \kappa_{\nu} \, \Sigma \, d\frac{\sigma
}{D^{2}}=\frac{2k<T>}{\lambda ^{2}D^{2}}\kappa_{\nu} M~~~~,
\end{equation}
where \( M=\int \, \Sigma d\sigma  \), $F_{\nu}$ is flux density,
$\kappa_{\nu}$ is the mass opacity, $\Sigma$ is the surface
density and $\sigma$ is the surface area of the disk
(\citealp{beckwith_s90}).  The dust temperatures T are almost
always sufficiently high (typically 50 K) to make the
Rayleigh-Jeans assumption valid, so the total emission is
proportional to a product of the total mass and the average
temperature $<T>$, weighted appropriately by its radial
distribution. The major uncertainty in mass estimates arises from
$\kappa_{\nu}$. Theoretically, $\kappa_{\nu}$ is expected to vary
as a power of the frequency $\nu$: $\kappa _{\nu }=\kappa
_{0}\left( \frac{\nu }{\nu _{0}}\right) ^{\beta }$. Following
\citet{beckwith_s90}, we adopt a fiducial value of $0.02\left(
\frac{\nu}{230 GHz}\right)$ cm$^2$g$^{-1}$ and calculate the total
gas mass assuming $M_{gas}/M_{dust}=100$. The disk mass of LkCa~15
is calculated to be $\sim$0.01 M$_{\odot}$.

\begin{figure}
\epsscale{0.65} \plotone{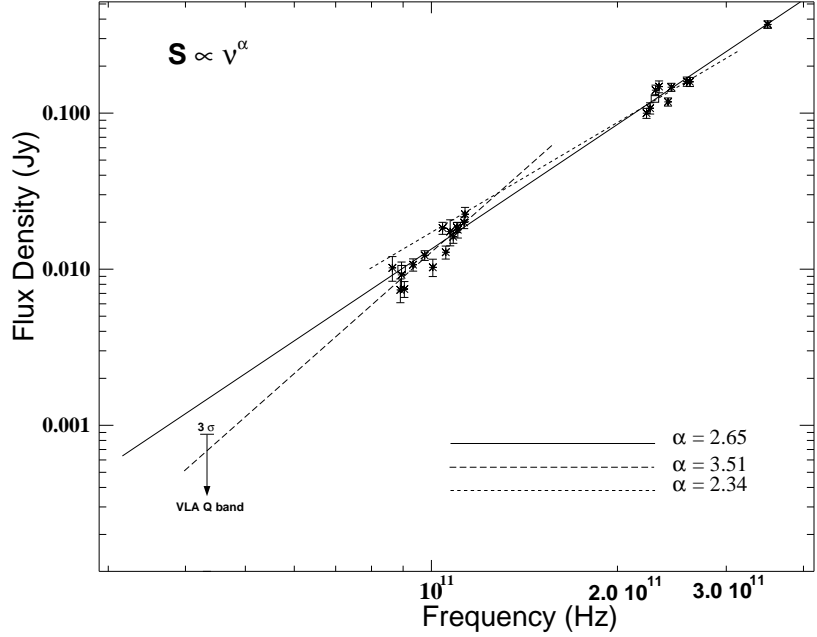} \caption{Measured
millimeter/submillimeter continuum emission fluxes toward LkCa~15.
The 850 $\mu$m point was acquired with SCUBA, whose image of
LkCa~15 is consistent with a point source (G. Sandell, private
communication). The two squares depict the 3 and 1.3 mm continuum
fluxes measured with the Plateau de Bure interferometer
(\citealp{duvert_g00}). The spectral slope as determined by all of
the available data is shown by the solid line, individual spectral
slope fits to the 3 mm and 1.3 mm data are depicted with the two
dashed lines. 7 mm VLA observations (D. Wilner, private
communication) detect no continuum emission to an rms of 0.29 mJy.
Even the 3$\sigma$ upper limit is inconsistent with an
extrapolation of the spectral index derived from the
millimeter/submillimeter fluxes.
 \label{fig:continuum}}
\end{figure}

\begin{figure}
\epsscale{1.0} \plotone{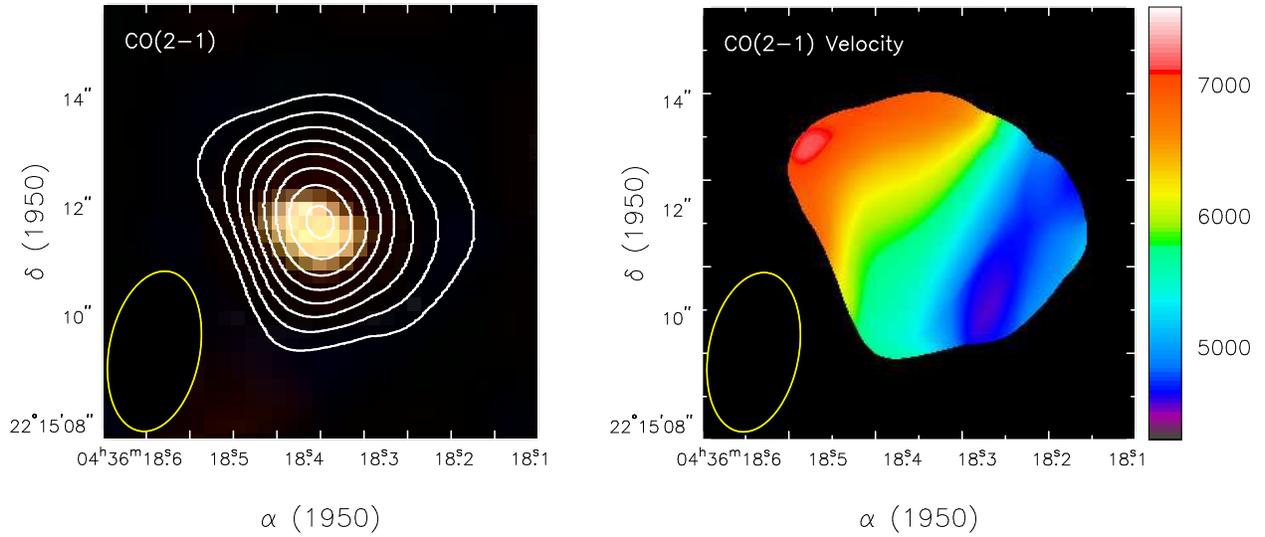} \caption{CO 2$\rightarrow$1
emission toward LkCa~15. The integrated intensity is shown in
contours at left, overlayed on the highest resolution continuum
image in gray scale.  Contours start at 30\%~of the peak value of
2.66 Jy km s$^{-1}$ and increase 10\%~thereafter.  The
intensity-weighted mean velocity field over the observed emission
region is shown at right ($v_{LSR}$ units in m s$^{-1}$). The
synthesized beam is shown at lower left.
 \label{fig:co}}
\end{figure}

\begin{figure}
\epsscale{1.}
\plotone{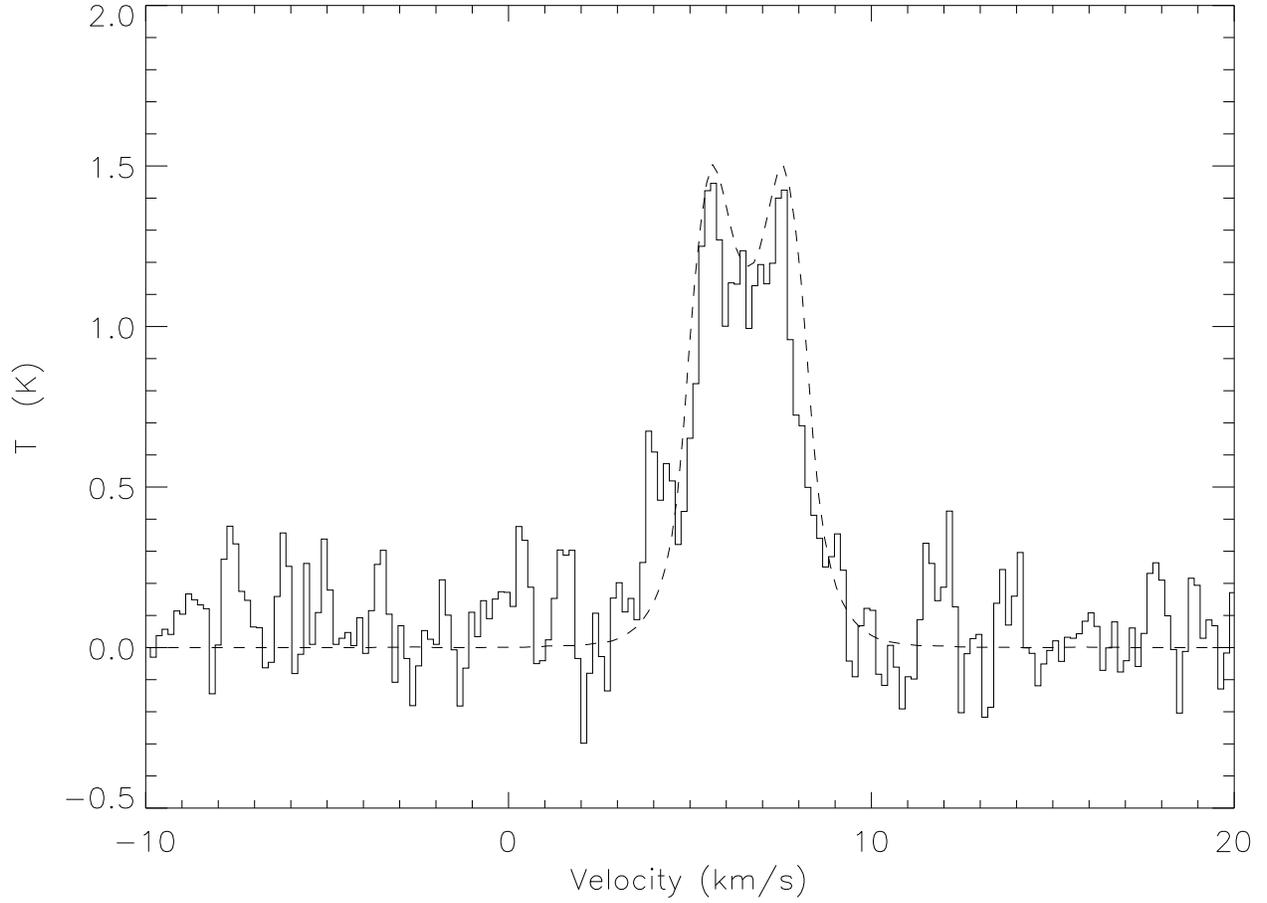}
\caption{The disk averaged spectrum of CO 2$\rightarrow$1 (solid histogram),
compared with the model CO emission for a 430 AU radius disk
inclined at 58$^{\circ}$ (dotted line). Details of the calculation
may be found in \S 3.2.
 \label{fig:co_spectra}}
\end{figure}

\begin{figure}
\epsscale{0.9} \plotone{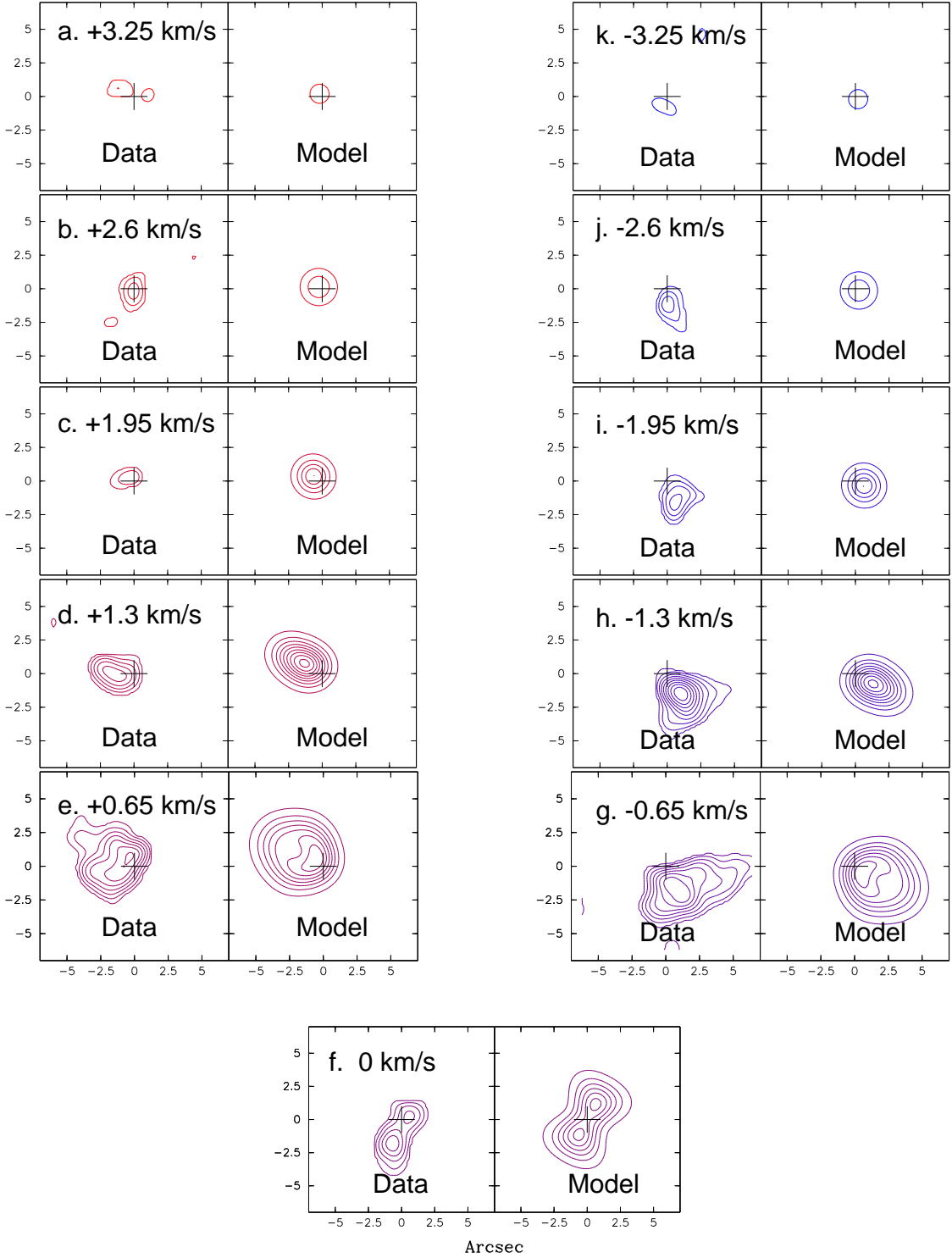} \caption{Spectral-line maps of the
CO 2$\rightarrow$1 emission toward LkCa~15, obtained with OVRO in
steps of 0.65 km s$^{-1}$, shown adjacent to simulations of the
emission predicted by a kinematic model of a disk in Keplerian
rotation with parameters as described in the text. The velocities
are refered to the stellar(systemic) velocity of $V_{lsr}=6.3$ km
s$^{-1}$.
 \label{fig:cochan}}
\end{figure}

\begin{figure}
\epsscale{1.}
\plotone{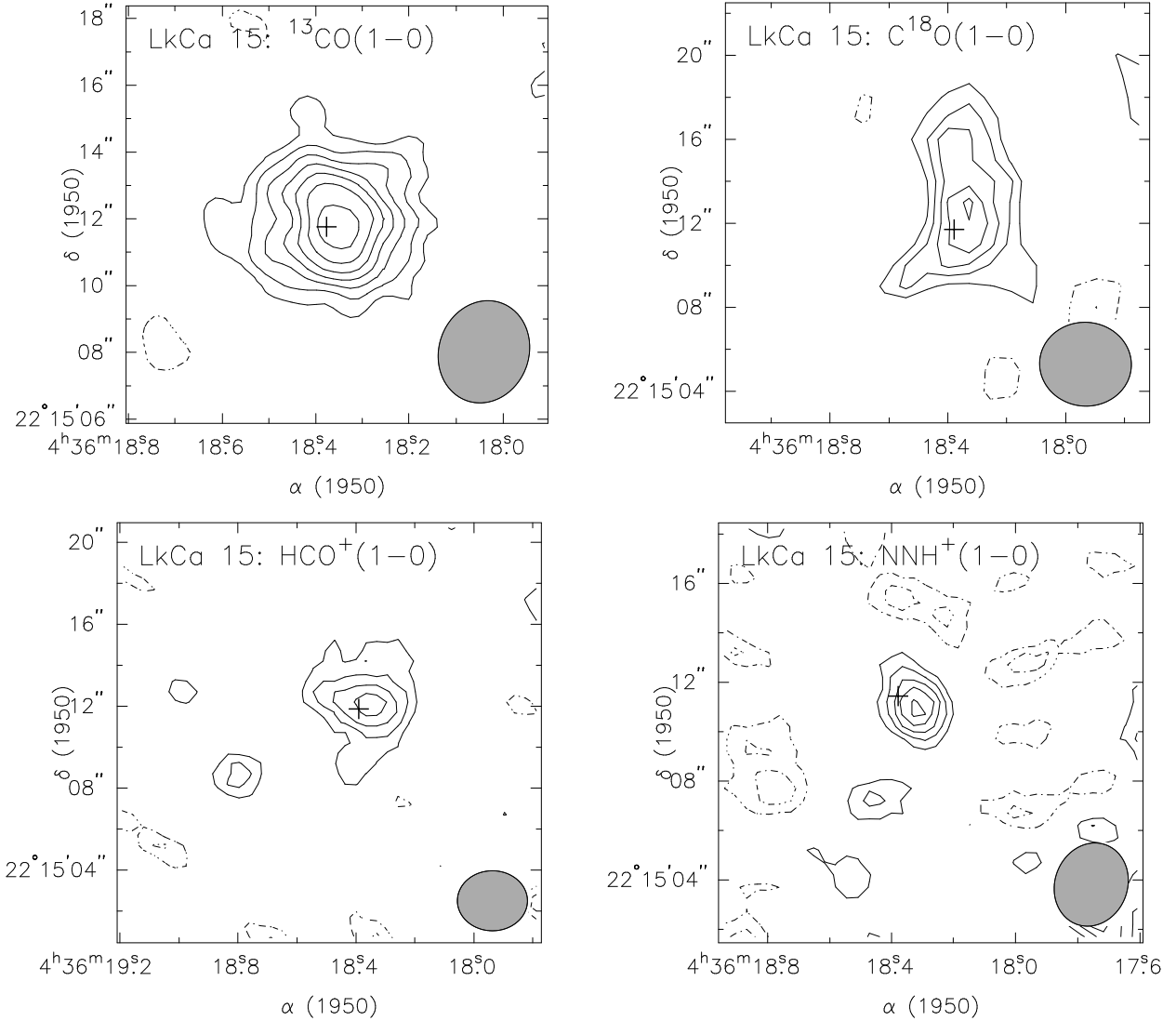}
\caption{Top: The integrated intensity maps of $^{13}$CO (left,
contours start at 0.16 Jy beam$^{-1}$ km s$^{-1}$ and are spaced
by 0.08 Jy beam$^{-1}$ km s$^{-1}$) and C$^{18}$O (right, contours
start at 0.1 Jy beam$^{-1}$ km s$^{-1}$ and are spaced by 0.05
Jy beam$^{-1}$ km s$^{-1}$) 1 $\rightarrow$ 0 emission.
Bottom: The integrated intensity maps of HCO$^+$ (left, contours
start at 0.16 Jy beam$^{-1}$ km s$^{-1}$ and are spaced by 0.08
Jy beam$^{-1}$ km s$^{-1}$) and N$_2$H$^+$ (right, contours
start at 0.1 Jy beam$^{-1}$ km s$^{-1}$ and are spaced by 0.05
Jy beam$^{-1}$ km s$^{-1}$) 1 $\rightarrow$ 0 emission. All steps
correspond to the rms noise level in each map.
 \label{fig:molecules}}
\end{figure}

\begin{figure}
\epsscale{1.}
\plotone{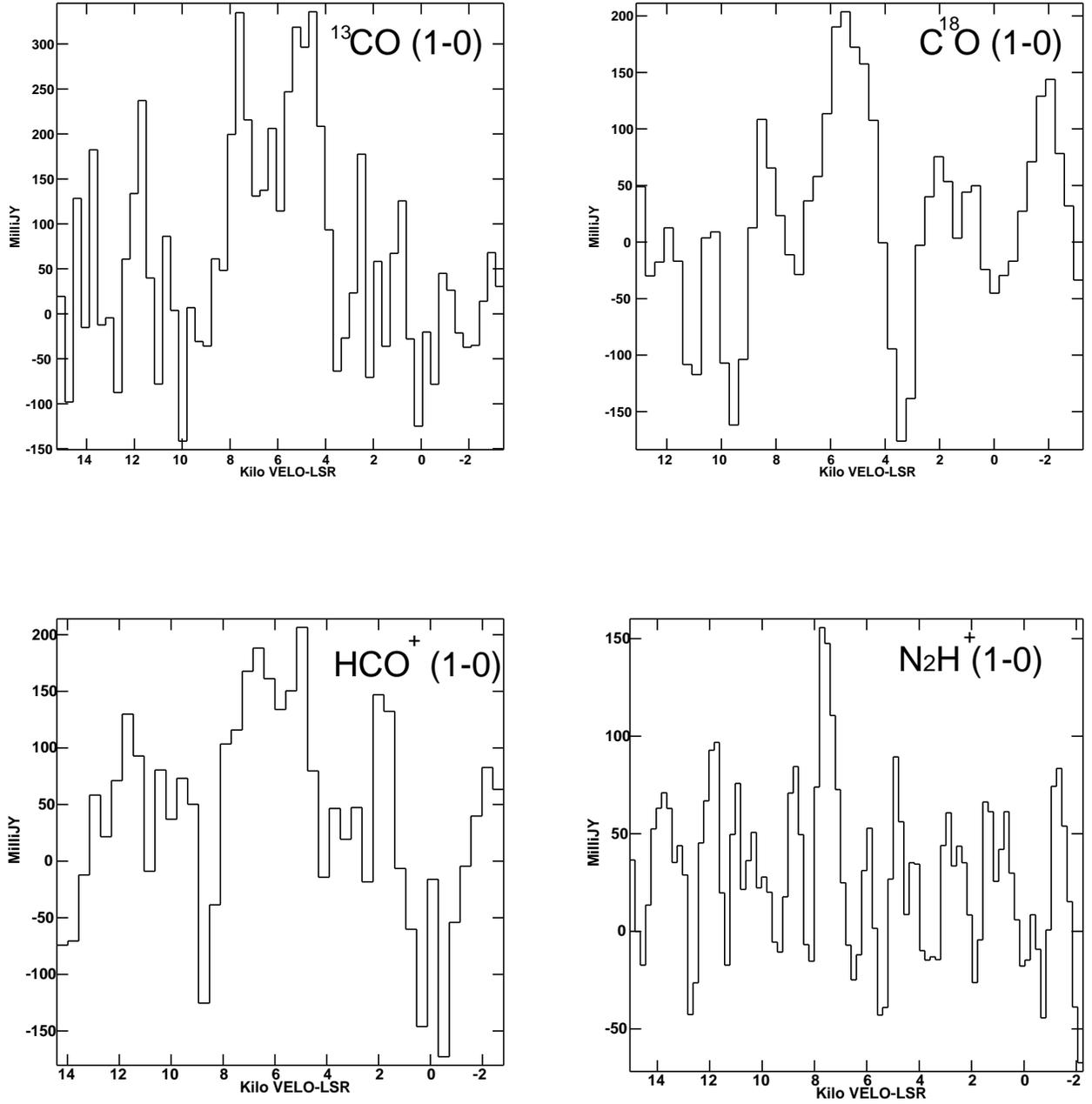}
\caption{Disk averaged spectra of the molecules in
Figure~\ref{fig:molecules}, obtained over boxes of 6$''$$\times$7$''$
for $^{13}$CO, 8$''$$\times$5$''$ for C$^{18}$O, 6$''$$\times$7$''$
for HCO$^+$, and 6$''$$\times$6$''$ for N$_2$H$^+$.
 \label{fig:molecules_spectra}}
\end{figure}

More detailed constraints for the density and temperature
structure can be derived by fitting the spectral energy
distributions (SED) of young star/disk systems.
Figure~\ref{fig:continuum} shows the flux density as a
function of frequency of the millimeter wavelength emission
from LkCa~15.
Because the continuum and molecular lines were observed
simultaneously, the frequencies of the continuum observations
at 1.3 and 2.7 mm vary slightly depending on the local oscillator
frequency required for the molecular line observations.  In
Figure~\ref{fig:continuum}, the spectral index, $\alpha$, as
defined by $F_{\nu} \approx {\nu}^{\alpha}$, is 2.65, consistent
with IRAM PdBI observations (\citealp{duvert_g00}).
A separate analysis of the data in each atmospheric window, however,
suggests that the slope of the 3 mm points alone ($\alpha=3.51$) is
considerably steeper than that for the 1.3 mm fluxes ($\alpha=2.34$),
although the error bars are large.

In order to examine this issue further, we compare our OVRO
3.4 -- 1.2 mm continuum fluxes with those obtained
at 870 $\mu$m and 7 mm by the JCMT and VLA, respectively. The
JCMT SCUBA flux lies along a spectral index $\alpha$ of 2.65
with respect to the OVRO data, and so is consistent with the
local slope derived from the OVRO 1.3 mm continuum emission
measurements. No simple extrapolation of the sub-millimeter
data can explain the VLA non-detection at 7 mm (D. Wilner,
private communication), which requires $\alpha$ values similar
to or even larger than those measured at 3 mm.
The coincidence of the 1.3 and 3 mm data observed by
\citet{duvert_g00} with the fits to our OVRO data indicate
that the observed change in the spectral index at millimeter
wavelengths is not likely to be due to errors in  flux calibration.

Several physical mechanisms, notably optical depth effects and
grain growth, can contribute to the observed behavior. As
mentioned above, ${\kappa}_{\nu}$ is expected to vary as
${\nu}^{\beta}$, where ${\beta}$ is the opacity index. Assuming
optically thin emission in the Rayleigh-Jeans limit, the opacity
index $\beta$ (=$\alpha-2$) varies from 1.51 to 0.34 for the LkCa
15 disk, while $\beta$ is approximately 2 for dust in the diffuse
interstellar medium.

How reasonable is this assumption of optically thin continuum
emission? Following \citet{beckwith_s90}:
\begin{equation}
\alpha -2=\frac{\beta }{1+\frac{p}{(2-q)\ln (\frac{2}{(2-p)\tau })}}
\end{equation}
where $\frac{p}{(2-q)\ln (\frac{2}{(2-p)\tau })}$ is the ratio of optically
thick to optically thin emission and depends on
the disk structure,
$ \tau _{\nu }=\frac{\kappa _{\nu }M_{D}}{\cos \theta \pi R_{D}^{2}} $ is
the average optical depth at frequency $\nu$, and $M_D$ is the mass
of the disk. Given the dust disk radius of 190 AU and an $M_D$ of
0.01 M$_{\odot}$ (derived above from F$_{\nu}$),
the average optical depth at 230 GHz is {\lapprox}0.03. Thus, for any
reasonable disk density and temperature distribution, the optical depth
correction does not appear to be sufficient to explain the observed
change in the LkCa~15 spectral index with a single $\beta$.

One means to solve this dilemma is for the mass opacity
coefficient $\kappa _{\nu}$ to decrease by nearly an order of
magnitude between 1.2 and 3.4 mm.  Recent detailed models of
irradiated T Tauri disks, including dust grain growth
(\citealp{dalessio_c01}), conclude that when grains grow from
$\mu$m to mm sizes, the mass opacity coefficient ${\kappa}_{\nu}$
changes significantly between 1.3 and 3 mm.  Indeed, observations
of circumstellar disks (\citealp{beckwith_s90,
mannings_e94,wilner_h96}) indicate typical values of beta around
0.3, similar to that expected from mm-sized pebbles ($0<\beta<2$),
as opposed to the steeper frequency dependence of ${\kappa}_{\nu}$
($\beta=2$) found in the general interstellar medium
(\citealp{hildebrand83}).

As indicated by \citealp{chiang_j01}, the disk mass and grain size
distribution cannot be uniquely constrained by the continuum SED
alone when only unresolved fluxes are available. Additionally,
\citet{dalessio_c01} find that for a given total dust mass, grain
growth is accompanied by a decrease in the vertical height of the
disk surface. Fits of the SED for LkCa~15 (\citealp{chiang_j01})
indicate that the ratio of the height of the disk photosphere (H)
to the vertical gas scale height (h) is small (1.0), which can be
interpreted as evidence that the dust has begun to settle
vertically toward the midplane. The H/h parameter is a rather
robust one within the confines of the model, for as H/h decreases
from 4 to 1 there is an corresponding decrease in the overall
level of infrared excess at ${\lambda} \leq$100 $\mu$m since the
disk intercepts less stellar radiation. Dust settling can help
explain the relative paucity of large disks to which the powerful
technique of scattered light coronography can be applied and
specifically the lack of any detectable reflection nebulosity in HST
WFPC2GTO observations of LkCa~15 (\citealp{krist_b97}). Thus, the
most likely explanation for the observed change in the spectral
index from 1 mm to 3 mm is a combination of grain growth and dust
settling.

\subsection{Radiative Transfer and Molecular Line Emission}

The observed line intensities in circumstellar disks are a complex function of
the physical structure of the disk and of the line/continuum optical depth. If
we assume that circumstellar disk abundances are similar to those found in
dark clouds, the emission lines from parent isotopologues of common species
such as CO, HCO$^+$, HCN, etc., should be highly optically thick. Our observed
ratios of $^{12}$CO/$^{13}$CO emission reveal that $^{12}$CO is indeed
optically thick, but say little about $^{13}$CO.  Several LkCa~15 transits were
therefore devoted to a local oscillator setting containing the C$^{18}$O J=1
$\rightarrow$ 0 transition, the results of which are  presented in
Figure~\ref{fig:molecules}. The observed beam-matched flux density ratio is 5,
only somewhat lower than the isotope ratio [$^{13}$C]/[$^{18}$O] of 8.3.  The
$^{13}$CO J=1 $\rightarrow$ 0 transition is therefore only somewhat optically
thick, and that of C$^{18}$O is optically thin, and should therefore provide a
more reliable estimate of the disk mass, {\em provided} it remains in the gas
phase. \citet{vanzadelhoff_v01} indicate that,  for standard HCO$^+$
abundances,  the 1$\rightarrow$0 line is again optically thick in the outer
layers, whereas that of H$^{13}$CO$^+$ is close to being optically thin
throughout the disk. Further, the low-J HCO$^+$ lines are predicted to probe
densities of order $10^5 - 10^6$ cm$^{-3}$, below the critical density of the
4$\rightarrow$3 transition (so that LTE cannot be safely assumed). For
H$^{13}$CO$^+$, the populations will be closer to thermal equilibrium because
its emission arises primarily from regions with densities of $10^7-10^8$
cm$^{-3}$ (\citealp{vanzadelhoff_v01}).  Parent isotopologue lines are
therefore likely to be optically thick for many species in disks, and so
whenever possible the emission from isotopically substituted species should
be utilized in deriving relative molecular abundances.

The observations described above indicate the need for detailed calculation
 of disk radiative transfer.  Furthermore, the complex geometry within the
disk demands that flexible and efficient radiative transfer techniques
be utilized, and of the many techniques available the Monte Carlo
approach has shown great promise (\citealp{choi_e95}).
Since Monte Carlo techniques utilize integration paths chosen at random,
any number of coordinate systems can be used. The high optical depth
of many transitions in circumstellar disks generates very short photon
mean free paths, however, and so unless aggressive acceleration schemes
are employed, the Monte Carlo approach can be very slow. Computational
instabilities and poor convergence can result, but efficient
one-dimensional (1D) and two-dimensional (2D) Monte Carlo treatments
of the line radiative transfer based on the two-layer passive disk
models outlined above have now begun to appear
(\citealp{hogerheijde_vdtak00,vanzadelhoff_v01}). In all such approaches,
the intensity of each line is calculated by solving the radiative transfer
equation.

2D implenetations are necessary to quantitatively treat inclined
disks. For this reason we have used an accelerated 2D Monte Carlo
model (\citealp{hogerheijde_vdtak00}) to examine the radiative
transfer and molecular excitation in the LkCa~15 disk, taking both
collisional and radiative processes into account. This model
produces a simulated observation of each transition as imaged by a
telescope with infinite resolution for a disk of a given size,
inclination and temperature distribution.  The MIRIAD function
UVMODEL and the observed visibility data set were then used to
sample this model at the observed $(u,v)$ spacings and the model
data set was processed in a manner identical to that of the OVRO
data, thus allowing a direct comparison of the two. For these
model calculations, the temperature and hydrogen density as a
function of radius and height were acquired from a model similar
to those presented in D'Alessio et al. (2001), but calculated
specifically for the stellar parameters of LkCa~15 such that the
overall SED of the star+disk is reproduced. In this model, we
assume spherical, compact dust grains with a power law size
distribution with exponent 3.5, a maximum grain size of 1 mm and
grain abundances from Pollack et al. (1994).  The accretion rate
is assumed to be $1.0\times10^{-8}$ M$_{\odot}$ yr$^{-1}$, with
the inner radius of the disk set at 5 AU and the outer radius at
425~AU, which is consistent with that derived above for LkCa~15.
The inclination (58 degrees) and turbulent velocity width (0.1 km
s$^{-1}$) are found by fitting the observed CO 2$\rightarrow$1
line profile with the model spectrum, using
N$_{CO}$=10$^{-4}$N$_H$~\footnote{This model assumes that CO and
H$_2$ have similar radial and vertical distributions. Chemical
models (i.e. \citet{aikawa_v02}) indicate that this is not
strictly the case, and variations in the vertical distribution of
CO will be explored in future work.} as shown in
Figure~\ref{fig:co_spectra}. For each transition observed, a suite
of models with the parameters described above were created, but
with the molecular abundances relative to hydrogen varied until a
match with the observed integrated line intensity was obtained.
Integrated intensity is a useful quantity for comparison because
the column density is approximately constant as a function of
radius (\citealp{willacy_l00,aikawa_h99}) at the large radii ($>$
50 AU) to which we are sensitive with the OVRO millimeter array.

The resulting column densities are shown in Table~\ref{tab:mol}.
The column densities calculated using this non-LTE radiative
transfer model are 1-2 orders of magnitude larger than those
calculated in the \citet{willacy_l00} models of disk chemistry,
but consistent with those calculated by \citet{aikawa_v02} -- except
that N$_2$H$^+$ is an order of magnitude larger than the model prediction.
As indicated in the last column of Table~\ref{tab:mol}, for CO
2$\rightarrow$1, $^{13}$CO 1$\rightarrow$0,  HCO$^+$
1$\rightarrow$0 and N$_2$H$^+$ 1$\rightarrow$0 emission the column
densities are smaller when calculated using the LTE assumption.
This is due to the fact that the emission from these transitions
is optically thick (see discussion above).
The LTE calculations assume that the gas is
optically thin and that the entire disk is being probed by
observation of this line  emission and thus result in an
underestimation of the total amount of CO, HCO$^+$ and N$_2$H$^+$
present. In the case of the C$^{18}$O 1$\rightarrow$0 and
H$^{13}$CO$^+$ 1$\rightarrow$0 emission, which are believed to be
optically thin, and for the conditions in the near-surface regions
of disks beyond 50 AU, the column densities are overestimated when
thermal equilibrium is assumed.  This supports the conclusion that
conditions in the emitting region are not at LTE, and that both
radiative and collisional processes play a part in the excitation.
The population of higher energy rotational states is thus much
lower than would be the case if collisions were dominant; and the
assumption of thermal equilibrium therefore results in an
overestimation of the population of the upper states and thus in
the total column density, which at LTE is directly proportional to
the density of the upper state.

Both the results presented above and observations of high-J
transitions of CO, HCO$^+$, and HCN toward LkCa~15 by van
Zadelhoff el al. (2001) suggest that the molecular emission from
the major isotopologues is optically thick and that the inferred
densities in the region are not sufficient to thermalize all
transitions. The line ratios presented in van Zadelhoff el al.
(2001) further indicate temperatures and densities typically
associated with warm layers lying above the disk midplane.
Additionally, if we ignore depletion of gas phase species onto
grains, and assume cosmic fractional abundances for CO
(10$^{-5}$--10$^{-4}$ N$_H$) and hydrogen column densities of
10$^{23}$ cm$^{-2}$, the resulting CO column density is an order
of magnitude greater than that observed here (10$^{18}$
cm$^{-2}$). Thus, for any disk in which the gas and dust are well
mixed and there is no depletion, even species such as C$^{18}$O
are predicted to be optically thick at molecular cloud abundances.
The low column densities derived here and the high-J/low-J line
ratios therefore indicate that the disk midplane (heights below
several tens AU) is not being probed by the observed emission and
that the column densities presented in Table 1 are only applicable
to the warm layers of the disk.

\subsection{Ionization Balance}

The theory of the ionization balance in molecular clouds and the relevant
ion-molecule chemistry has been widely used for determinations of the
electron abundances in dense clouds (\citealp{langer85}). Theoretical chemical
models indicate that ion-molecule reactions are also efficient in the
outer regions  of protoplanetary disks (\citealp{aikawa_u98}).
Photoionization, cosmic ray ionization and, possibly, X-ray ionization
of H$_2$ rapidly lead to the formation of H$_3^+$, the pivotal species
in ion-molecule models of molecular cloud chemistry.  The low proton
affinity of H$_2$ ensures that H$_3^+$, which has no rotational spectrum,
will undergo proton transfer reactions with abundant neutral species such
as CO and N$_2$.  In disks, gaseous CO and N$_2$ exist in regions well
removed from the midplane, where the temperature warms enough for sublimation
of these molecules (T $\approx$ 20-30 K).  The isoelectronic ions
HCO$^+$ and N$_2$H$^+$ were observed here and can therefore be used to
estimate the fractional ionization in the warm layer of the LkCa~15 disk.

The volume density in the emitting region can be constrained from
line ratios obtained from the results presented here and previous
single dish observations of molecules with high dipole moments
(e.g. HCO$^+$, HCN).  An LVG analysis for LkCa~15, performed by
van Zadelhoff et al. (2001), indicates that the lines arise from
regions with $n_{\rm H_2}$$\sim$10$^5$--10$^7$ cm$^{-3}$. We use
10$^6$ cm$^{-3}$ as the number density of H$_2$ in the
calculations below, which is also the number density at a height
of $\sim$150 AU and a radius of 350 AU in the D'Alessio disk
model. The mean kinetic temperatures are less well constrained,
but the analysis of multiple transitions of CO (van Zadelhoff et
al. 2001) suggests that T$_{kin}$$\sim$20--40 K for LkCa~15. We
therefore adopt a single kinetic temperature of 30 K. Also, for
the reasons discussed in \S 3.2, we use the column densities
resulting from the non-LTE radiative transfer model summarized in
Table 1 for all observed species. Because H$_2$ cannot be observed
directly at millimeter wavelengths, the column density is highly
model dependent. Typical values vary from 10$^{23}$ to 10$^{24}$
cm$^{-2}$ for R$\sim$300 AU, the distance corresponding to the
typical beam size of our observations. van Zadelhoff et al. (2001)
investigate the level of molecular depletion by statistical
equilibrium calculations as compared to the gas column densities
predicted by dust continuum observations, and find depletion by a
factor of 3--40 for CO in the disk of LkCa~15. We therefore adopt
10$^{-5}$ as the fractional abundance of CO, by assuming depletion
by a factor of 10 from the nominal value of $f$(CO) in molecular
clouds. In our calculations we also assume that the warm layers in
the disk are in local (chemical) equilibrium and that the
abundances of CO, HCO$^+$ and N$_2$H$^+$ peak within these layers
(Aikawa et al. 2002), although the vertical distributions of these
species are not identical [cf. Figure 4 of
\citet{vanzadelhoff_a03} and Figures 4 and 5 of
\citet{willacy_l00}].

To simplify the calculations presented here, we first consider
only the ionization balance determined by HCO$^+$, H$_3^+$,
N$_2$H$^+$ and electrons in steady state, for which
\begin{equation}
k_{1}[H_{3}^{+}][CO]=k_{e}(HCO^{+})[HCO^{+}][e]-k_{2}[N_{2}H^{+}][CO]
\label{eq:hco}
\end{equation}

\begin{equation}
k_{e}(H_{3}^{+})[H_{3}^{+}][e]+k_{3}[N_{2}][H_{3}^{+}]+k_{1}[H_{3}^{+}][CO]=\frac{\zeta }{n(H_{2}
)}
\label{eq:h3}
\end{equation}

\begin{equation}
k_{3}[N_{2}][H_{3}^{+}]=k_{2}[N_{2}H^{+}][CO]+k_{e}(N_{2}H^{+})[N_{2}H^{+}][e]
\label{eq:n2h}
\end{equation}

\noindent
where $k_e(H_3^+)$ = 1.15 $\times$ 10$^{-7}$ $ \left( \frac{T}{300}\right) ^{-0.65}$ cm$^3$s$^{-1}$,
$k_{e}(HCO^+)$ = 2.0$\times$ 10$^{-7}$ $ \left( \frac{T}{300}\right) ^{-0.75}$
cm$^3$ s$^{-1}$ and $k_{e}(N_2H^+)$ = 1.7 $\times$ 10$^{-7}$ $ \left( \frac{T}{300}\right) ^{-0.9}$
cm$^3$ s$^{-1}$ are the dissociative recombination rate coefficients of
H$_3^+$, HCO$^+$ and N$_2$H$^+$,
k$_1$ = 6.56$\times$ 10$^{-10}$ $ \left( \frac{T}{300}\right) ^{-0.5}$ cm$^3$ s$^{-1}$ and
k$_2$ = 8.8 $\times$ 10$^{-10}$ cm$^3$ s$^{-1}$ are the reaction rate coefficients of CO with H$_3^+$
and N$_2$H$^+$, respectively, k$_3$ = 1.8 $\times$ 10$^{-9}$ is the rate coefficient for
the reaction $N_2 + H_3^+ \rightarrow N_2H^+ + H_2$
(all rate coefficients have been adopted from the UMIST database
for astrochemistry (\citealp{millar_f97})), and the bracket [ ]
represents the fractional abundance of the species and $\zeta$ the
ionization rate due to cosmic rays, X-rays, and UV radiation.

A lower limit to the electron fraction follows directly from the
observed HCO$^+$ and N$_2$H$^+$ column densities, but can also be
obtained more ``locally'' by noting that the right side of
equation~\ref{eq:hco} must be larger than 0, or:

\begin{equation}
\frac{[e]}{[CO]}\geq \frac{k_{2}[N_{2}H^{+}]}{k_{e}(HCO^{+})[HCO^{+}]}\sim 1 \times 10^{-3}
\label{eq:e1}
\end{equation}

\noindent The lower limit of the fractional ionization $[e]$ is
therefore {\grapprox}10$^{-8}$, given [CO]$\sim$10$^{-5}$, in the
near-surface region of the disk. This value is fairly high
compared to that of cold, dark clouds at $n_{\rm H_2}$$\sim$10$^6$
cm$^{-3}$ where for example, \citet{caselli_w02} obtained
fractional ionization of 10$^{-9}$ in the center of L1544, and we
note that the ionization fraction in LkCa~15 may be enhanced
relative to other T Tauri stars. The ROSAT All-Sky Survey of young
stars in Taurus (\citealp{neuhaeuser_s95}) shows that the
probability of the existence of an X-ray source for LkCa~15 is
98\% and the upper limit for its X-ray luminosity is 4.1$\times$
10$^{29}$ ergs$^{-1}$.  Ionization from the resulting high-energy
radiation may have important effects on the disk
(\citealp{feigelson_m99}). The high HCO$^+$ abundance and CN/HCN
ratio observed (Qi, Kessler \& Blake 2003) indicate that LkCa~15
could possess similar chemistry to the intense X-ray source TW
Hya, another Sun-like star with strong molecular emission from its
attendant disk (\citealp{kastner_z97}).

Understanding the ionization rate near the disk surface is
important because the electron fraction scales linearly with the
ionization rate.  A full 3D calculation of X-ray transport and
ionization in axially asymmetric disks has been developed by
\citet{igea_g99} using a Monte Carlo approach. They calculate the
ionization rate at $R=1$ AU in a minimum mass solar nebula disk
model with L$_x$ = 10$^{29}$ erg s$^{-1}$, which can be scaled to
other X-ray luminosities and disk radii and is found to be
relatively insensitive to changes in disk parameters. Scaling
their values at $R=300$ AU with N(H$_2$)$\sim$10$^{23}$ cm$^{-2}$,
we calculate an X-ray ionization rate of $\sim$10$^{-16}$ s$^{-1}$
near the surface of the LkCa~15 disk. This is nearly a factor of
ten larger than the galactic cosmic ray ionization rate, and shows
that X-rays can play an important role in the disk ionization
balance. The enhanced ionization can drive additional chemical
activity, and the inclusion of X-rays within the context of dust
settling models developed for LkCa~15 may well account for much
higher abundances of molecules observed toward this source as
compared with the DM Tau disk, for which ``standard'' disk
chemistry models produce good agreement with observations
(\citealp{aikawa_v02}).

By considering the balance of all the molecular ions present in
the disk, even tighter constraints can be placed on the electron
fraction. We consider two conditions with differing abundances of
metal ions, as discussed in \citet{langer85}.  When the abundances
of metal ions are low, H$_3^{+}$, HCO$^+$ and N$_2$H$^+$ become
the major cations within the disk. Because HCO$^+$ and N$_2$H$^+$
have fractional abundances that are only {\grapprox}10$^{-10}$
while the lower limit to the electron fraction is closer to
10$^{-8}$, H$_3^{+}$ dominates. In this case, the electron
abundance can be calculated simply from $[e]\simeq (\frac{\zeta
}{k_{e}n(H_{2})})^{\frac{1}{2}}\sim 1.4 \times 10^{-8}$
(\citealp{langer85}), which yields values similar to that derived
above. However, the high resulting abundance of H$_3^+$ drives
rapid protonation of CO, that is $k_{1}[H_{3}^{+}][CO]\gg
k_{e}(HCO^{+})[HCO^{+}][e]$, and makes HCO$^+$ balance
(equation~\ref{eq:hco}) impossible within the confines of the
simple model.

\begin{figure}
\epsscale{1.}
\plotone{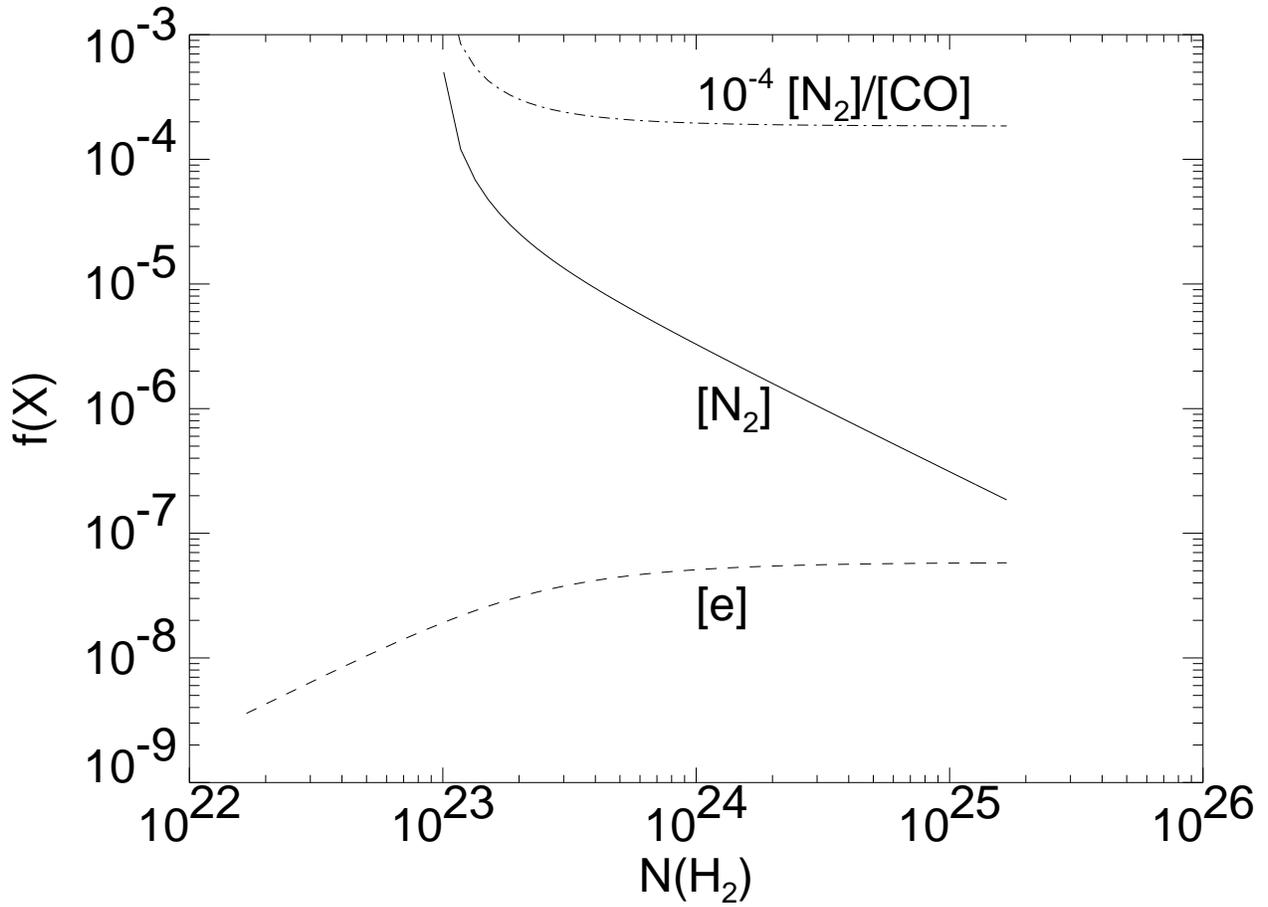}
\caption{Fractional abundances ($f(X)=N(X)/N(H_2)$ of N$_2$ and
electrons versus the H$_2$ column density for the simple fractional
ionization model discussed in \S 3.3.
 \label{fig:ions_model}}
\end{figure}

It is therefore reasonable to examine the contribution of metal
ions to the fractional ionization because they are destroyed very
slowly and can comprise a significant fraction of the ion
abundance even though their elemental abundances are small.
\citet{fromang_t02} calculate the ionization fraction of
protostellar disks, taking into account vertical temperature
structure and the possible presence of trace metal atoms, and find
that a tiny fraction of the cosmically available metals can
dramatically affect the ionization balance. Considering the model
above, but now with $[H_3^+]+[HCO^+]+[N_2H^+]{\ll}[e]$,
we can solve for the 3 unknown variables $[N_2]$, $[H_{3}^+]$ and $[e]$ in
order to balance equations~\ref{eq:hco},\ref{eq:h3},\ref{eq:n2h}.
Using equations~\ref{eq:h3} and \ref{eq:n2h}, removing the
term $k_{3}[N_{2}][H_{3}^{+}]$, and inserting the value of
H$_{3}^{+}$ derived from the equation~\ref{eq:hco}, the
equation for $[e]$ becomes:
\begin{equation}
k_{e}(HCO^{+})k_{e}(H_{3}^{+})[HCO^{+}][e]^{2}+B[e]-\frac{\zeta }{n(H_{2})}k_{1}[CO]=0
\label{eq:e}
\end{equation}
where $B=-k_{2}k_{e}(H_{3}^{+})[CO][N_{2}H^{+}]+k_{1}k_{e}(HCO^{+})[CO][HCO^{+}]+k_{1}k_{e}(N_{2}H^{+})[CO][N_{2}H^{+}]$.
This equation can be solved using the observed
column densities of CO, HCO$^+$, and N$_2$H$^+$ and an assumed
density of H$_2$. The N$_2$/CO ratio can also be derived
from equations~\ref{eq:n2h} and \ref{eq:e} under such conditions,
and Figure~\ref{fig:ions_model} presents the solutions for a range
of H$_2$ column densities. Not surprisingly, the electron fraction
is calculated to be larger than the metal free case ($\sim$10$^{-8}$),
 which, as we stress above, can only be
applied for the physical conditions prevailing near the
warm surface layer(s) of the disk.

Transport processes in the outer disk likely depend critically on
the degree of ionization. Efficient coupling between the neutral
material and magnetic fields is thought to be provided by
fractional ionizations of {\grapprox}$10^{-8}$
(\citealp{feigelson_m99}). As we have outlined above, photons,
X-rays and cosmic rays can be important sources of ionization near
the disk surface.  The photon penetration depth is very short
compared to the disk thickness, however, and even cosmic rays are
exponentially attenuated for columns $>$100 g cm$^{-2}$. The
electron fraction can therefore be expected to drop toward the
midplane due to the increasing density, suggesting the possibility
of layered accretion (\citealp{gammie96}), with a turbulent
surface layer. However, due to the depletion of volatiles such as
CO, the disk mid-plane fractional ionization may be larger than
expected from a direct extrapolation of the electron abundance
inferred for the surface layer. For very high depletions, electron
recombination begins to control the fractional abundance of
H$_3^+$, resulting in a higher than expected electron (and
H$_3^+$) abundance. The details of the vertical and radial
ionization fraction are pivotal to magnetohydrodynamic (MHD)
models of YSO accretion disk viscous transport mechanisms
(\citealp{glassgold_f00}). \citet{balbus_h91}, for example,
propose that differential rotation drives instabilities in the
perpendicular magnetic field of accretion disks. Such
magnetorotational instabilities then create a turbulent disk if
they are well coupled to the neutral disk material. As shown
above, observations suggest that the combination of photons,
X-rays, and cosmic rays can produce the ionization needed to
couple the magnetic field to the neutral disk material near the
disk surface. Chemically, transport or mixing processes between
the surface layer and the disk interior would be necessary to
bring freshly made molecules such as CN, HCN or even deuterated
species back to the interior of the disk. In this scenario, the
extent of such processes, which are dependent on the ionization
fraction, would greatly effect the composition of early solar
system objects such as comets.

\subsection{CO Depletion and the N$_2$/CO Ratio}

If CO remains primarily in the gas phase, it is sufficiently
chemically stable that its abundance will be little altered from
dark cloud values. CO would therefore be a good tracer of the disk
gas mass, but depletion onto grain mantles in the cold disk is
likely to substantially reduce the {\em gas phase} column density
of CO and many other species. Thus, estimates of the degree of
depletion of volatile molecules such as CO and N$_2$ onto grains
will greatly improve our understanding of disk chemistry. For CO,
depletion factors can be measured directly from infrared
spectroscopy of edge on disks (\citealp{boogert02}), or more
generally by using observations of the dust continuum along with
several isotopically-substituted molecules (e.g. $^{12}$CO,
$^{13}$CO, C$^{18}$O) to distinguish depletion from optical depth
effects. From the column density of CO ($1.7\times 10^{18}$
cm$^{-2}$) obtained with the detailed 2D Monte Carlo radiative
transfer model (Table 1), a disk gas mass of only $\sim$10$^{-4}$
M$_{\odot}$ is derived, a value in agreement with that derived
from single dish observations of $^{13}$CO higher-J transitions
(\citealp{thi01}) using the JCMT and CSO but nearly two orders of
magnitude less than the mass derived from dust emission -- far
more than can be accounted for by errors either in the dust mass
opacity coefficient or in the spectral line radiative transfer.
Depletion thus appears to be a dominant effect and while CO and
its isotopomers are therefore excellent tracers of the disk
velocity field, it is clear that they are not reliable tracers of
the disk mass. In the models of Aikawa and co-workers
(\citealp{aikawa_m96, aikawa_u99}), CO depletes onto grains at
radii greater than 200-300 AU. Inside this radius, the temperature
warms to the sublimation temperature of CO (T $\sim$ 20 K) and the
gas phase column density rises substantially. If thermal
sublimation is the main desorption mechanism, the depletion
patterns in disks should be strongly correlated with volatility.
We have observed strong emission from non-volatile species such as
HCN and CN toward LkCa~15 with spatial distributions that are not
well correlated with CO (Qi, Kessler \& Blake 2003). Additional
mechanisms are therefore likely to be important in maintaining the
observed gas-phase abundances.

As a closed shell homonuclear diatomic molecule, N$_2$ has no electric
dipole allowed rotational transitions and so cannot be observed directly
at millimeter wavelengths.  However, from our N$_2$H$^+$ observations
and the model presented above,
we can estimate the N$_{2}$/CO ratio, and find it to be
$\sim$2 (Figure~\ref{fig:ions_model}).
This is much higher than the ratios inferred for typical
dark clouds at moderate densities, which are $\sim$0.01--0.1
(\citealp{womack_z92, bergin_c01}). In the dark cloud IC5136,
\citet{bergin_c01} do find evidence for the presence of differential
gas-phase depletion and altered CO/N$_2$ ratios in the densest
portions, where CO exhibits a
significant abundance reduction and N$_2$H$^+$ is relatively
undepleted -- although in their models the abundance of N$_2$ is
still less than that of CO.

Within disks, \citet{aikawa_h99} investigated the
two-dimensional chemical structure within the so-called Kyoto
minimum mass solar nebula model (\citealp{hayashi81}), and
predicted much lower abundances of
N$_2$ than CO at all vertical scales. Because they adopted an
artificially low sticking probability to reproduce the spectra
of CO available at that time, without specifying any non-thermal
desorption processes or modifying the temperature distribution in
the Kyoto model, such an approach might not account for the known
differential desorption of CO and N$_2$. Further, the vertical
temperature gradients of actual protoplanetary disks should be
more complex than those in the Koyoto model (\citealp{cg97, dalessio_c97}).
\citet{willacy_l00} investigated the molecules in the super-heated layer
of the Chiang and Goldreich (1997) disk models, and found that
molecules in this layer are destroyed by the harsh ultraviolet
radiation from the star, although in the warm upper layers of the D'Alessio
et al. models UV shielding is sufficient to prevent destruction of most
molecules (\citealp{aikawa_v02}).
We therefore believe that N$_2$ and its chemical product
N$_2$H$^+$ trace dense gas in the disk where CO is largely depleted but
N$_2$ is not, and that this region is protected by an overlying
warm layer that protects N$_2$ from ultraviolet radiation,
in order to produce the high N$_2$/CO ratio observed.

Deriving absolute fractional abundances is more difficult, but
from the N$_2$/CO ratio and our Monte Carlo radiative transfer
calculations we estimate a fractional abundance for N$_2$ of
$\sim$ 2 $\times$ 10$^{-5}$. N$_2$ would therefore represent a
substantial fraction of the cosmically available nitrogen, this
plus the gas phase N$_2$/CO ratio inferred for LkCa 15 provides a
direct observational connection to comets. The abundance of N$_2$
in comets is an important guide to their volatile content because
it is trapped and released by amorphous ice in a manner similar to
that of Ar (\citealp{bar-nun_k88}). This process has implications
for understanding the role of comets for delivery of volatiles and
noble gases to the terrestrial planets. \citet{wyckoff_t91} have
reviewed the total nitrogen abundance in comet Halley and argued
that it is depleted by 2-6 fold relative to that of the sun. The
laboratory experiments performed by \citet{bar-nun_k88} showed
that CO is trapped 20 times more efficiently than N$_2$ in
amorphous ice formed at 50 K and \citet{owen_b95} predicted that
in icy planetesimals forming in the solar nebula at about 50 K,
N$_2$/CO $\approx$ 0.06 in the gases trapped in the ice provided
N$_2$/CO $\approx$ 1 in the nebula itself, a value quite
consistent with our data. \citet{wyckoff_t89} reported N$_2$/CO
$\sim$ $2 \times 10^{-3}$ in Halley's comet, whereas
\citet{lutz_w93} found N$_2$/CO $\sim$ $2 \times 10^{-2}$ in Comet
Bradfield, and N$_2$/CO $\sim$ $3 \times 10^{-2}$ in Comet
Kohoutek. The observations are actually of the two molecular ions
(CO$^+$ and N$_2 ^+$) and the conversion from the ion abundances
to those of the neutrals is dependent on poorly constrained
photodestruction branching ratios. \citet{wyckoff_t89} argued that
the ion ratio must be multiplied by a factor of 2, while
\citet{lutz_w93} find no factor necessary. Large amounts of
gaseous N$_2$ may thus be required in the outer solar nebula, as
we find in the disk of LkCa~15, in order to explain the N$_2$/CO
ratios found in comets. This can be confirmed by measurements of
the N$_2$/CO ratio in dynamically new comets, which should have
higher values than those in short-period comets
(\citealp{owen_b95}).

\section{Summary}

We have detected molecular gas and continuum emission from the
disk encircling the T Tauri star LkCa~15 at frequencies between 85
and 260 GHz. The 1.2 mm dust continuum emission is resolved in our
$1.''35\times0.''74$ beam with a minimum diameter of 190 AU and an
inclination angle of $\sim$57$^\circ$. A noticeable decrease in
the continuum spectral slope with frequency may result from the
combination of grain growth and dust settling. Sensitive
observations at longer wavelengths and submillimeter continuum
images are needed to confirm the trend. Based on Keplerian
rotation model fits to the CO velocity field, the gas emission
extends to $\sim$750 AU while the characteristic radius of the
disk is determined to be closer to 425 AU from Gaussian fits to
the CO 2$\rightarrow$1 spectral data cube.

We have also presented molecular images of various isotopologues
of CO and the HCO$^+$ and N$_2$H$^+$ ions in the disk. Our
observations show that most of the molecular emission from the
dominant species of common molecules is optically thick. Different
molecules and even different transitions of one molecule will
therefore probe different layers of the disk; most of the emitting
region detected is near the disk surface. Detailed radiative
transfer modelling indicates that LTE is not appropriate in these
regions. With this 2D model and the column densities from the
detection of the HCO$^+$ and N$_2$H$^+$ ions, we derive a lower
limit to the fractional ionization of 10$^{-8}$ and an N$_2$/CO
ratio of 2, which reveals a region where CO is heavily depleted
but N$_2$ is not. This zone must be well protected by the
efficient shielding of ultraviolet radiation by an overlying disk
layer. Additional discussions concerning the UV field and its
impact on disk chemistry will be presented in forthcoming papers.

The Owens Valley Radio Observatory is supported by NSF grant AST-9981546.
We are grateful to those OVRO staff members who diligently
scheduled these time-consuming observations. We thank P. D'Alessio
for providing the disk model of LkCa~15 and D. Wilner, G. Sandell for
sharing their unpublished data.
J.E.K. is supported by the NASA Graduate Student Researchers Program,
NGT5-50231. G.A.B. gratefully acknowledges support from the NASA
Exobiology and Origins of Solar Systems programs.

\clearpage

\begin{deluxetable}{lccccc}
\tablewidth{0pt}
\tablecaption{Observed molecular intensities and column densities toward LkCa~15 \label{tab:mol}}
\tablehead{
\colhead{Transition}&\colhead{Beam}&\colhead{$\int T dv$} &\colhead{N(30K,LTE)} &\colhead{N(Model)} &\colhead{Model Ratios}\\
\colhead{} &\colhead{(arcsec)}&\colhead{(K km s$^{-1}$)} &\colhead{(cm$^{-2}$)} &\colhead{(cm$^{-2}$)} &\colhead{N(Model)/N(30K,LTE)} }
\startdata
CO 2$\rightarrow$1            &  1.81$\times$1.49 &  $~$12.5 &  $~$7.28(15) &  1.68(18)    &  230  \\
$^{13}$CO 1$\rightarrow$0     &  2.74$\times$2.41 &  $~$6.39 &  $~$1.11(16) &  3.04(16)    &  2.74 \\
C$^{18}$O 1$\rightarrow$0     &  5.05$\times$3.59 &  $~$1.90 &  $~$3.31(15) &  1.40(15)    &  0.42 \\
HCO$^+$ 1$\rightarrow$0       &  7.74$\times$5.64 &  $~$3.30 &  $~$9.25(12) &  2.31(13)    &  1.79 \\
H$^{13}$CO$^+$ 1$\rightarrow$0&  6.30$\times$4.41 &  $<$0.88 &  $<$2.60(12) &  $<$1.12(12) &  0.43 \\
N$_2$H$^+$ 1$\rightarrow$0    &  3.51$\times$3.17 &  $~$3.83 &  $~$1.71(13) &  3.07(13)    &  1.80 \\
\enddata
\end{deluxetable}


\clearpage

\begin{thebibliography}{55}
\expandafter\ifx\csname natexlab\endcsname\relax\def\natexlab#1{#1}\fi

\bibitem[{{Aikawa} \& {Herbst}(1999)}]{aikawa_h99}
{Aikawa}, Y. \& {Herbst}, E. 1999, \apj, 526, 314

\bibitem[{{Aikawa} \& {Herbst}(2001)}]{aikawa_h01}
---. 2001, \aap, 371, 1107

\bibitem[{{Aikawa} {et~al.}(1996){Aikawa}, {Miyama}, {Nakano}, \&
  {Umebayashi}}]{aikawa_m96}
{Aikawa}, Y., {Miyama}, S.~M., {Nakano}, T., \& {Umebayashi}, T. 1996, \apj,
  467, 684

\bibitem[{{Aikawa} {et~al.}(1998){Aikawa}, {Umebayashi}, {Nakano}, \&
  {Miyama}}]{aikawa_u98}
{Aikawa}, Y., {Umebayashi}, T., {Nakano}, T., \& {Miyama}, S. 1998, in
  Chemistry and Physics of Molecules and Grains in Space. Faraday Discussions
  No. 109, 281

\bibitem[{{Aikawa} {et~al.}(1999){Aikawa}, {Umebayashi}, {Nakano}, \&
  {Miyama}}]{aikawa_u99}
{Aikawa}, Y., {Umebayashi}, T., {Nakano}, T., \& {Miyama}, S.~M. 1999, \apj,
  519, 705

\bibitem[{{Aikawa} {et~al.}(2002){Aikawa}, {van Zadelhoff}, {van Dishoeck}, \&
  {Herbst}}]{aikawa_v02}
{Aikawa}, Y., {van Zadelhoff}, G.~J., {van Dishoeck}, E.~F., \& {Herbst}, E.
  2002, \aap, 386, 622

\bibitem[{{Alexander} {et~al.}(2001){Alexander}, {Boss}, \&
  {Carlson}}]{alexander_b01}
{Alexander}, C.~M.~O., {Boss}, A.~P., \& {Carlson}, R.~W. 2001, Science, 293,
  64

\bibitem[{{Amelin} {et~al.}(2002){Amelin}, {Krot}, {Hutcheon}, \&
  {Ulyanov}}]{amelin_k02}
{Amelin}, Y., {Krot}, A., {Hutcheon}, I., \& {Ulyanov}, A. 2002, Science, 297,
  1678

\bibitem[{{Balbus} \& {Hawley}(1991)}]{balbus_h91}
{Balbus}, S.~A. \& {Hawley}, J.~F. 1991, \apj, 376, 214

\bibitem[{{Bar-Nun} {et~al.}(1988){Bar-Nun}, {Kleinfeld}, \&
  {Kochavi}}]{bar-nun_k88}
{Bar-Nun}, A., {Kleinfeld}, I., \& {Kochavi}, E. 1988, \prb, 38, 7749

\bibitem[{{Beckwith} {et~al.}(1990){Beckwith}, {Sargent}, {Chini}, \&
  {Guesten}}]{beckwith_s90}
{Beckwith}, S. V.~W., {Sargent}, A.~I., {Chini}, R.~S., \& {Guesten}, R. 1990,
  \aj, 99, 924

\bibitem[{{Bergin} {et~al.}(2001){Bergin}, {Ciardi}, {Lada}, {Alves}, \&
  {Lada}}]{bergin_c01}
{Bergin}, E.~A., {Ciardi}, D.~R., {Lada}, C.~J., {Alves}, J., \& {Lada}, E.~A.
  2001, \apj, 557, 209

\bibitem[{{Boogert} {et~al.}(2002){Boogert}, {Hogerheijde} \& {Blake}}]{boogert02}
{Boogert}, A.~C.~A., {Hogerheijde}, M.~R., \& {Blake}, G.~A. 2002,
\apj, 568, 761

\bibitem[{{Bouvier} {et~al.}(1995){Bouvier}, {Covino}, {Kovo}, {Martin},
  {Matthews}, {Terranegra}, \& {Beck}}]{bouvier_c95}
{Bouvier}, J., {Covino}, E., {Kovo}, O., {Martin}, E.~L., {Matthews}, J.~M.,
  {Terranegra}, L., \& {Beck}, S.~C. 1995, \aap, 299, 89

\bibitem[{{Caselli} {et~al.}(2002){Caselli}, {Walmsley}, {Zucconi}, {Tafalla},
  {Dore}, \& {Myers}}]{caselli_w02}
{Caselli}, P., {Walmsley}, C.~M., {Zucconi}, A., {Tafalla}, M., {Dore}, L., \&
  {Myers}, P.~C. 2002, \apj, 565, 344

\bibitem[{{Chiang} \& {Goldreich}(1997)}]{cg97}
{Chiang}, E.~I. \& {Goldreich}, P. 1997, \apj, 490, 368

\bibitem[{{Chiang} {et~al.}(2001){Chiang}, {Joung}, {Creech-Eakman}, {Qi},
  {Kessler}, {Blake}, \& {van Dishoeck}}]{chiang_j01}
{Chiang}, E.~I., {Joung}, M.~K., {Creech-Eakman}, M.~J., {Qi}, C., {Kessler},
  J.~E., {Blake}, G.~A., \& {van Dishoeck}, E.~F. 2001, \apj, 547, 1077

\bibitem[{{Choi} {et~al.}(1995){Choi}, {Evans}, {Gregersen}, \&
  {Wang}}]{choi_e95}
{Choi}, M., {Evans}, N.~J., {Gregersen}, E.~M., \& {Wang}, Y. 1995, \apj, 448,
  742

\bibitem[{{D'Alessio} {et~al.}(1997){D'Alessio}, {Calvet}, \&
  {Hartmann}}]{dalessio_c97}
{D'Alessio}, P., {Calvet}, N., \& {Hartmann}, L. 1997, \apj, 474, 397

\bibitem[{{D'Alessio} {et~al.}(2001){D'Alessio}, {Calvet}, \&
  {Hartmann}}]{dalessio_c01}
---. 2001, \apj, 553, 321

\bibitem[{{Duvert} {et~al.}(2000){Duvert}, {Guilloteau}, {M{\'e}nard}, {Simon},
  \& {Dutrey}}]{duvert_g00}
{Duvert}, G., {Guilloteau}, S., {M{\'e}nard}, F., {Simon}, M., \& {Dutrey}, A.
  2000, \aap, 355, 165

\bibitem[{{Elias}(1978)}]{elias78}
{Elias}, J.~H. 1978, \apj, 224, 857

\bibitem[{{Feigelson} \& {Montmerle}(1999)}]{feigelson_m99}
{Feigelson}, E.~D. \& {Montmerle}, T. 1999, \araa, 37, 363

\bibitem[{{Fromang} {et~al.}(2002){Fromang}, {Terquem}, \&
  {Balbus}}]{fromang_t02}
{Fromang}, S.~., {Terquem}, C., \& {Balbus}, S.~A. 2002, \mnras, 329, 18

\bibitem[{{Gammie}(1996)}]{gammie96}
{Gammie}, C.~F. 1996, \apj, 457, 355

\bibitem[{{Glassgold} {et~al.}(2000){Glassgold}, {Feigelson}, \&
  {Montmerle}}]{glassgold_f00}
{Glassgold}, A.~E., {Feigelson}, E.~D., \& {Montmerle}, T. 2000, Protostars and
  Planets IV, 429

\bibitem[{{Hayashi}(1981)}]{hayashi81}
{Hayashi}, C. 1981, Progress of Theoretical Physics, 70, 35

\bibitem[{{Hildebrand}(1983)}]{hildebrand83}
{Hildebrand}, R.~H. 1983, \qjras, 24, 267

\bibitem[{{Hogerheijde} \& {van der Tak}(2000)}]{hogerheijde_vdtak00}
{Hogerheijde}, M.~R. \& {van der Tak}, F. F.~S. 2000, \aap, 362, 697

\bibitem[{{Horne} \& {Marsh}(1986)}]{horne_m86}
{Horne}, K. \& {Marsh}, T.~R. 1986, \mnras, 218, 761

\bibitem[{{Igea} \& {Glassgold}(1999)}]{igea_g99}
{Igea}, J. \& {Glassgold}, A.~E. 1999, \apj, 518, 848

\bibitem[{{Kastner} {et~al.}(1997){Kastner}, {Zuckerman}, {Weintraub}, \&
  {Forveille}}]{kastner_z97}
{Kastner}, J.~H., {Zuckerman}, B., {Weintraub}, D.~A., \& {Forveille}, T. 1997,
  Science, 277, 67

\bibitem[{{Kessler}, {Qi} \& {Blake}(2003)}]{kessler03}
{Kessler}, J.~E., {Qi}, C. \& {Blake}, G.~A. 2003, \apj, in preparation

\bibitem[{{Kitamura} {et~al.}(2002){Kitamura}, {Momose}, {Yokogawa}, {Kawabe},
  {Tamura}, \& {Ida}}]{kitamura_m02}
{Kitamura}, Y., {Momose}, M., {Yokogawa}, S., {Kawabe}, R., {Tamura}, M., \&
  {Ida}, S. 2002, \apj, 581, 357

\bibitem[{{Koerner}(1995)}]{koerner95}
{Koerner}, D.~W. 1995, Ph.D.~Thesis, California Institute of Technology

\bibitem[{{Krist} {et~al.}(1997){Krist}, {Burrows}, {Stapelfeldt}, \& {WFPC2 Id
  Team}}]{krist_b97}
{Krist}, J.~E., {Burrows}, C.~J., {Stapelfeldt}, K.~R., \& {WFPC2 Id Team}.
  1997, in American Astronomical Society Meeting, Vol. 191, 0514

\bibitem[{{Lada}(1999)}]{lada99}
{Lada}, E.~A. 1999, in NATO ASIC Proc. 540: The Origin of Stars and Planetary
  Systems, 441

\bibitem[{{Langer}(1985)}]{langer85}
{Langer}, W.~D. 1985, in Protostars and Planets II, 650--667

\bibitem[{{Lay} {et~al.}(1997){Lay}, {Carlstrom}, \& {Hills}}]{lay_c97}
{Lay}, O.~P., {Carlstrom}, J.~E., \& {Hills}, R.~E. 1997, \apj, 489, 917

\bibitem[{{Lutz} {et~al.}(1993){Lutz}, {Womack}, \& {Wagner}}]{lutz_w93}
{Lutz}, B.~L., {Womack}, M., \& {Wagner}, R.~M. 1993, \apj, 407, 402

\bibitem[{{Mannings} \& {Emerson}(1994)}]{mannings_e94}
{Mannings}, V. \& {Emerson}, J.~P. 1994, \mnras, 267, 361

\bibitem[{{Mannings} {et~al.}(1997){Mannings}, {Koerner}, \&
  {Sargent}}]{mannings_k97}
{Mannings}, V., {Koerner}, D.~W., \& {Sargent}, A.~I. 1997, \nat, 388, 555

\bibitem[{{Millar} {et~al.}(1997){Millar}, {Farquhar}, \&
  {Willacy}}]{millar_f97}
{Millar}, T.~J., {Farquhar}, P.~R.~A., \& {Willacy}, K. 1997, \aaps, 121, 139

\bibitem[{{Neuhaeuser} {et~al.}(1995){Neuhaeuser}, {Sterzik}, {Schmitt},
  {Wichmann}, \& {Krautter}}]{neuhaeuser_s95}
{Neuhaeuser}, R., {Sterzik}, M.~F., {Schmitt}, J. H. M.~M., {Wichmann}, R., \&
  {Krautter}, J. 1995, \aap, 297, 391

\bibitem[{{Owen} \& {Bar-Nun}(1995)}]{owen_b95}
{Owen}, T. \& {Bar-Nun}, A. 1995, Icarus, 116, 215

\bibitem[{{Qi}, {Kessler} \& {Blake}(2003)}]{qi03}
{Qi}, C. {Kessler}, J.~E., \& {Blake}, G.~A. 2003, \apj, in preparation

\bibitem[{{Robberto} {et~al.}(1999){Robberto}, {Meyer}, {Natta}, \&
  {Beckwith}}]{robberto_m99}
{Robberto}, M., {Meyer}, M.~R., {Natta}, A., \& {Beckwith}, S. V.~W. 1999, ESA
  SP-427: The Universe as Seen by ISO, 427, 195

\bibitem[{{Sargent} \& {Beckwith}(1987)}]{sargent_b87}
{Sargent}, A.~I. \& {Beckwith}, S. 1987, \apj, 323, 294

\bibitem[{{Scoville} {et~al.}(1993){Scoville}, {Carlstrom}, {Chandler},
  {Phillips}, {Scott}, {Tilanus}, \& {Wang}}]{scoville_c93}
{Scoville}, N.~Z., {Carlstrom}, J.~E., {Chandler}, C.~J., {Phillips}, J.~A.,
  {Scott}, S.~L., {Tilanus}, R. P.~J., \& {Wang}, Z. 1993, Publ.. Astron. Soc.
  Pac., 105, 1482

\bibitem[{{Simon} {et~al.}(2001){Simon}, {Dutrey}, \& {Guilloteau}}]{simon_d01}
{Simon}, M., {Dutrey}, A., \& {Guilloteau}, S. 2001, \apj, 545, 1034

\bibitem[{{Skrutskie} {et~al.}(1990){Skrutskie}, {Dutkevitch}, {Strom},
  {Edwards}, {Strom}, \& {Shure}}]{skrutskie90}
{Skrutskie}, M.~F., {Dutkevitch}, D., {Strom}, S.~E., {Edwards}, S., {Strom},
  K.~M., \& {Shure}, M.~A. 1990, \aj, 99, 1187

\bibitem[{{Strom} {et~al.}(1989){Strom}, {Strom}, {Edwards}, {Cabrit}, \&
  {Skrutskie}}]{strom_s89}
{Strom}, K.~M., {Strom}, S.~E., {Edwards}, S., {Cabrit}, S., \& {Skrutskie},
  M.~F. 1989, \aj, 97, 1451


\bibitem[{{Thi} {et~al.}(2001){Thi}, {van Dishoeck}, {Blake}, {van Zadelhoff},
  {Horn}, {Becklin}, {Mannings}, {Sargent}, {van den Ancker}, {Natta}, \&
  {Kessler}}]{thi01}
{Thi}, W.~F., {van Dishoeck}, E.~F., {Blake}, G.~A., {van Zadelhoff}, G.~J.,
  {Horn}, J., {Becklin}, E.~E., {Mannings}, V., {Sargent}, A.~I., {van den
  Ancker}, M.~E., {Natta}, A., \& {Kessler}, J. 2001, \apj, 561, 1074

\bibitem[{{van Zadelhoff} {et~al.}(2003){van Zadelhoff}, {Aikawa},
  {Hogerheijde}, \& {van Dishoeck}}]{vanzadelhoff_a03}
{van Zadelhoff}, G.-J., {Aikawa}, Y., {Hogerheijde}, M.~R., \& {van Dishoeck},
  E.~F. 2003, \aap, 397, 789

\bibitem[{{van Zadelhoff} {et~al.}(2001){van Zadelhoff}, {van Dishoeck}, {Thi},
  \& {Blake}}]{vanzadelhoff_v01}
{van Zadelhoff}, G.-J., {van Dishoeck}, E.~F., {Thi}, W.-F., \& {Blake}, G.~A.
  2001, \aap, 377, 566

\bibitem[{{Willacy} \& {Langer}(2000)}]{willacy_l00}
{Willacy}, K. \& {Langer}, W.~D. 2000, \apj, 544, 903

\bibitem[{{Wilner} {et~al.}(1996){Wilner}, {Ho}, \& {Rodriguez}}]{wilner_h96}
{Wilner}, D.~J., {Ho}, P.~T.~P., \& {Rodriguez}, L.~F. 1996, \apjl, 470, L117

\bibitem[{{Womack} {et~al.}(1992){Womack}, {Ziurys}, \& {Wyckoff}}]{womack_z92}
{Womack}, M., {Ziurys}, L.~M., \& {Wyckoff}, S. 1992, \apj, 393, 188

\bibitem[{{Wyckoff} {et~al.}(1991){Wyckoff}, {Tegler}, \&
  {Engel}}]{wyckoff_t91}
{Wyckoff}, S., {Tegler}, S.~C., \& {Engel}, L. 1991, \apj, 367, 641

\bibitem[{{Wyckoff} \& {Theobald}(1989)}]{wyckoff_t89}
{Wyckoff}, S. \& {Theobald}, J. 1989, Advances in Space Research, 9, 157

\end{thebibliography}
\end{document}